\documentclass[11pt]{amsart}
\usepackage{wrapfig}
\usepackage{graphicx}

\newcommand{\noi}{\noindent}

\newcommand{\fr}{\frac}
\newcommand{\hb}{\hbar}

\newcommand{\vp}{{\bf p}}
\newcommand{\vP}{{\bf P}}
\newcommand{\vk}{{\bf k}}
\newcommand{\vr}{{\bf r}}
\newcommand{\ak}{{b_\vk}}
\newcommand{\aka}{{b^+_\vk}}
\newcommand{\vm}{{\bf m}}
\newcommand{\am}{{b_\vm}}
\newcommand{\ama}{{b^+_\vm}}

\newcommand{\om}{\omega}

\newcommand{\ran}{\rangle}
\newcommand{\lan}{\langle}

\newcommand{\del}{\delta}

\newcommand{\beg}{\begin{equation}}
\newcommand{\bg}{\begin{equation}}
\newcommand{\en}{\end{equation}}

\newcommand{\ba}{\begin{eqnarray}}
\newcommand{\bac}{\begin{array}{c}}
\newcommand{\eac}{\end{array}}
\newcommand{\bal}{\begin{eqnarray}{lcl}}
\newcommand{\eal}{\end{eqnarray}}
\newcommand{\ea}{\end{eqnarray}}

\newcommand{\dsum}{\displaystyle\sum}

\newcommand{\nn}{\nonumber}
\newcommand{\bt}{\begin{tabular}}
\newcommand{\et}{\end{tabular}}

\begin{document}

\thispagestyle{empty}
\begin{center}
\null\vspace{-1cm}

\medskip

\vspace{2cm}
{\bf  IMPROVABLE UPPER BOUNDS \\
TO THE  PIEZOELECTRIC  POLARON GROUND STATE ENERGY}\\
\vspace{1.8cm}
 A.V. Soldatov\footnote{soldatov@mi.ras.ru}\\
{\it V.A.Steklov Mathematical Institute of the Russian Academy of
Sciences,\\ 8 Gubkina Str., 119991 Moscow, Russia.}
\end{center}
\vspace{1cm}
\centerline{\bf Abstract}
\baselineskip=18pt
\bigskip

It was shown that an infinite sequence of improving non-increasing
upper bounds to the ground state energy (GSE) of a slow-moving
piezoeletric  polaron  can be devised.

\noi {\bf Key words:} piezoelectric polaron, ground state energy,
upper bound, variational method

\noi {\bf PACS:}  71.38.-k, 71.38.Fp

\newpage

\section{The piezoelectric polaron model}

 A quantized polaron model for the case of an electron interacting
with acoustic phonons through piezoelectric deformation potential
was introduced by A.R. Hutson \cite{piezotens}, its Hamiltonian
structure being similar to the Fr\"ohlich optical polaron model
introduced by H. Fr\"ohlich \cite{Fr3} earlier:

\beg H= \fr{\hat \vp^2}{2m}+\dsum_\vk \hb\om_{\vk}\aka\ak
+\dsum_\vk \tilde V_k\left(\aka e^{-i\vk\hat\vr}+  \ak
e^{i\vk\hat\vr}\right), \label{original}\en

\noi where $\om_{\vk}=sk$ is the frequency of the acoustical
phonons with $s$ being the velocity of sound,

\[ \tilde V_k=\left(\fr{4\pi
\alpha}{\tilde V}\right)^{1/2}\fr{\hb^2}{m}k^{-1/2},
\]

\noi where $\tilde V$ is the volume of the crystal, and

\[ \alpha=\fr{1}{2}e^2\fr{\langle e^2_{ijk}\rangle}{\varepsilon C s\hb}\]

\noi is the dimensionless coupling constant where $\langle
e^2_{ijk}\rangle$ is an average of the piezoelectric tensor
\cite{piezotens}, $\varepsilon$ is the dielectric constant and $C$
is an average elastic constant. The operators $\hat\vp$ and
$\hat\vr$ are the electron momentum and position coordinate
quantum operators,

\[
\quad [\hat p_i, \hat r_j]=-i\hb\del_{ij},
\]

\noi and the Bose operators $b^+_{\vk}$, $b_{\vk}$,

\[ [b_{\vk}, b_{\vk'}^+]=\del_{\vk\vk'},\quad [b_{\vk}, b_{\vk'}]=0,
\]

\noi create and annihilate phonons of one effective acoustic mode
 of energy $\hbar\om_{\vk}$ and wave vector $\vk$.

In what follows the energies will be expressed in units of
$2ms^2$, the length in units of $\hb/2ms$ and the phonon wave
vectors in units of $2ms/\hb$ so that all variables are
dimensionless. In this units the model (\ref{original}) reads as

\beg H= \hat \vp^2+\dsum_\vk k \aka\ak +\dsum_\vk V_k\left(\aka
e^{-i\vk\hat\vr}+  \ak e^{i\vk\hat\vr}\right),
\label{originaldl}\en

\noi with

\[  V_k=\left(\fr{4\pi
\alpha}{ V}\right)^{1/2}k^{-1/2},
\]

\noi where $V$ is dimensionless volume. Ultimately,  the sum over
the phonon vectors $\sum_{\vk}$ is to be replaced by the integral
$V/(2\pi)^3 \int d\vk$ with a finite cutoff at $k_0$, the boundary
of the first Brillouin zone in the phonon wave vector space, which
is introduced to account for the discreteness of the crystal
lattice with  $k_0 \sim 1/a$, where $a$ is the lattice constant.

\section{Piezoelectric polaron GSE}

As usual, one can introduce the polaron total momentum

\[
\hat \vP=\hat\vp+\dsum_{\vk}\vk \aka\ak
\]

\noi being a constant of the motion and commuting with the
Hamiltonian (\ref{original}). Hence, the Hamiltonian can be
transformed to the representation in which $\hat \vP$ becomes a
"c"-number $P$, the value of the total polaron momentum,

\[
H\to\tilde H,\quad \tilde H= S^{-1}HS, \quad
S=\exp(-i\dsum_{\vk}\vk\hat\vr\aka\ak),
\]

\beg \tilde H= (\vP - \dsum_{\vk}\vk \aka\ak )^2+\dsum_\vk
k\aka\ak +\dsum_\vk V_k\left(\aka + \ak \right),
\label{corigtransf}\en

\noi and the Hamiltonian (\ref{corigtransf}) does not contain the
electron coordinates anymore. One more subsequent transformation

\[
\tilde H\to {\mathcal H}(f),\quad {\mathcal H}(f)= U^{-1}\tilde
HU, \quad U=\exp\{\dsum_{\vk} f_{\vk}(\aka  - \ak )  \},
\]

\noi results in the Hamiltonian

\bg {\mathcal H}(f)={\mathcal H}_0(f)+{\mathcal H}_1(f),
\label{finalHam}\en

\noi where

\[ {\mathcal H}_0(f)= P^2+\dsum_{\vk} k\aka\ak+(\dsum_{\vk} \vk
\aka\ak)^2 -\alpha', \]

\ba {\mathcal
H}_1(f)=\dsum_{\vk}[(k+k^2)f_{\vk}+V_k](\aka+\ak)+2\dsum_{\vk\vm}(\vk\cdot
{\bf m} ) f_{\vk}f_{\vm} \aka\am + \nn \\
+\dsum_{\vk\vm}(\vk\cdot {\bf m} ) f_{\vk}f_{\vm} (\aka\ama
+\ak\am) +
2\dsum_{\vk\vm}(\vk\cdot {\bf m} ) f_{\vk}(\ama\am\ak+\aka\ama\am  )-\nn\\
-2\dsum_{\vk}(\vP\cdot {\bf k} )(\aka+f_{\vk})(\ak+f_{\vk})+\nn\\
+2\dsum_{\vk\vm} (\vk\cdot\vm) f_{\vm}^2\aka\ak + 2\dsum_{\vk\vm}
(\vk\cdot\vm) f_{\vm}^2(\aka+\ak)+\dsum_{\vk\vm} (\vk\cdot\vm)
f_{\vm}^2f_{\vk}^2, \label{altfinalHam} \ea

 \noi
 and

 \[
-\alpha'=
2\dsum_{\vk}V_kf_{\vk}+\dsum_{\vk}(k+k^2)f^2_{\vk}+(\dsum_{\vk}f^2_{\vk}\vk)^2.
 \]

\section{On improvable upper bounds to various polaron models GSE in general}

Our objective here is to demonstrate that an infinite, in
principle, sequence of improvable upper bounds to the ground state
energy $E(\alpha, \vP, k_0)$ of the Hamiltonian (\ref{finalHam})
can be constructed in a regular way by means of a variational
method outlined in \cite{SolVar1}. Similar variational approach
was formulated later in \cite{Kir1}. This energy corresponds to
the lowest energy of the slow-moving polaron for a given value of
the total polaron momentum $\vP$.  Of course, some reservations
are to be made here regarding the fact that, unlike in the case of
optical polaron, there is no energy separation between the polaron
ground state and the excited states with the same total momentum.
Therefore, what is considered here is actually a zero temperature
polaron behavior.

 Then, expansion of the function $E_g(\alpha,\vP, k_0)$ in powers of $\vP$
 of the kind

\[ E_g(\alpha, \vP, k_0)=E_g(\alpha,0,k_0) + \fr{P^2}{2m_{eff}} +
O(P\,^4), \]

\noi where $E_g(\alpha,0,k_0)$ is the GSE  of the polaron at rest,
may provide us with some approximate value of the polaron
effective mass $m_{eff}$ in spatially isotropic as well as
anisotropic case, in which general case the so-called inverse
effective mass tensor $\left( \fr{1}{m_{eff}}\right)_{ij}$  is to
be introduced as

\[ \left( \fr{1}{m_{eff}}\right)_{ij}=\fr{\partial^2 E(\alpha,\vP,
k_0)}{\partial P_i\partial P_j}\Bigg|_{\vP=0}. \]

\noi instead of the scalar effective mass parameter $m_{eff}$.


As was shown in \cite{SolVar1}, for a quantum system Hamiltonian
$\hat H$ and a trial state $|\psi\rangle$,\, such that
$\langle\psi|\psi\rangle =1$,

\[ E_g \le \min(a_1^{(n)},...,a_n^{(n)})\le \langle\psi| \hat
H|\psi\rangle, \]

\noi where the  real numbers $(a_1^{(n)},...,a_n^{(n)})$ are the
roots of the n-th order polynomial equation

\[ P_n(x)=\dsum_{i=0}^n X_ix^{n-i}=0. \]

\noi Here, the coefficient $X_0\equiv 1$ and all the other
coefficients $X_i$,\, $1\le i\le n$ satisfy the system of $n$
linear equations

\[ {\mathcal M} {\bf X} +{\bf Y} =0, \]

\noi with

\[ Y_i=M_{2n-i}, \quad {\mathcal M}_{ij}=M_{2n-(i+j)}, \quad i,j
=1,2,...n, \]

\noi and

\[ M_m = \langle\psi| \hat H^m|\psi\rangle.\]

\noi It is assumed that all moments $M_m$ are finite. It was
proved that  the following inequality holds

\[ \min(a_1^{(n+1)},...,a_{n+1}^{(n+1)}) \le
\min(a_1^{(n)},...,a_n^{(n)}). \]

\noi So, at the first order

\bg E_g\le a_1^{(1)},\qquad  a_1^{(1)}= \langle\psi| \hat
H|\psi\rangle, \label{firstorder}\en

\noi and at the second order

\bg E_g\le \min(a_1^{(2)}, a_2^{(2)}) = \langle\psi| \hat
H|\psi\rangle+\fr{K_3}{2K_2}-\left[
\left(\fr{K_3}{2K_2}\right)^2+K_2 \right]^{1/2},\label{secorder}
\en

\[ a_1^{(2)} = \langle\psi| \hat
H|\psi\rangle+\fr{K_3}{2K_2}-\left[
\left(\fr{K_3}{2K_2}\right)^2+K_2 \right]^{1/2}, \]

\[  a_2^{(2)} = \langle\psi| \hat
H|\psi\rangle+\fr{K_3}{2K_2}+\left[
\left(\fr{K_3}{2K_2}\right)^2+K_2 \right]^{1/2}, \]

\noi where  $K_2$ and $K_3$ are the central moments

\[ K_2= \langle\psi| (\hat H -  \langle\psi| \hat H|\psi\rangle
)^2|\psi\rangle, \qquad K_3= \langle\psi| (\hat H - \langle\psi|
\hat H|\psi\rangle )^3|\psi\rangle. \]

\noi It is seen  that the second order  bound (\ref{secorder})
would lie  below the first order  bound (\ref{firstorder}) for
most physically relevant quantum models and most reasonable and
meaningful choices of the trial state $|\psi\rangle$.

\section{Improvable upper bounds for piezoelectric polaron at rest}

 For $\vP=0$, the function $f_{\vk}$ is  spherically symmetric, and the piezoelectric polaron
 modelHamiltonian
 (\ref{finalHam}) reduces to

\ba {\mathcal H}(f)=  \dsum_{\vk} k\aka\ak+(\dsum_{\vk} \vk
\aka\ak)^2
-\alpha'+\nn \\
+ \dsum_{\vk}[(k+k^2)f_k+V_k](\aka+\ak)+2\dsum_{\vk\vm}(\vk\cdot
{\bf m} ) f_kf_m \aka\am + \nn \\
+\dsum_{\vk\vm}(\vk\cdot {\bf m} ) f_kf_m (\aka\ama +\ak\am) +
2\dsum_{\vk\vm}(\vk\cdot {\bf m} ) f_k(\ama\am\ak+\aka\ama\am
).\nn
 \ea

 \noi As to be seen later on, it would be convenient to choose phonon vacuum state $|0\rangle$ as a trial state $|\psi\ran$ for ${\mathcal
 H}(f)$,\, so that inequality

 \[ E_g(\alpha,0,k_0)\le \langle 0|{\mathcal H}(f)|0\rangle =2\dsum_{\vk}V_kf_k+\dsum_{\vk}(k+k^2)f^2_k
 \]

\noi holds, the right-hand side of which is minimized by

\[ f_k=-V_k/(k+k^2), \]

\noi and, eventually,

\bg  E_{g}(\alpha,0,k_0)\le E_{W}(\alpha,0,k_0)=
-\fr{2\alpha}{\pi}\ln[1+k_0].\label{LLP} \en

\noi The upper bound (\ref{LLP}) stems from the conventional
variational principle in quantum mechanics and is valid for
arbitrary value of $\alpha$. To derive a sequence of better
non-increasing upper bounds one needs to calculate moments $\lan
0|{\mathcal H}^m(f)|0\ran $ for sufficiently large integer
exponents $m$ by means of the Wick theorem. The result of the
calculation for any such moment can be expressed in terms of the
products of integrals of rational functions

\bg \int_0^{k_0}\fr{k^p dk}{(k+k^2)^q}, \qquad p,q -
\mbox{non-negative integers},\label{exint}\en

\noi which can be evaluated  analytically.

Thus, at the second order variational approximation
(\ref{secorder})

\ba E_g(\alpha,0,k_0)\le E_{var}= -\fr{2\alpha}{\pi}\ln[1+k_0]
+\fr{K_3}{2K_2}-\left[ \left(\fr{K_3}{2K_2}\right)^2+K_2
\right]^{1/2}, \label{SO}\ea

\noi where

\ba K_2= \fr{8\alpha^2}{3\pi^2} F_1^2(k_0), \label{K2}\ea

\ba K_3=\fr{16\alpha^2}{3\pi^2}
(F_1(k_0)F_2(k_0)+F_1(k_0)F_3(k_0)) + \fr{64\alpha^3}{9\pi^3}
F_1^3(k_0), \label{K3}
 \ea

\ba F_1(k_0)= \left[\ln(1+k_0)+\fr{1}{1+k_0}-1 \right], \label{F1}
\ea

\ba F_2(k_0)= \left[\fr{k_0^2}{2}-k_0+\ln(1+k_0) \right],
\label{F2} \ea

\ba F_3(k_0)=
\left[-2k_0+\fr{k_0^2}{2}+3\ln(1+k_0)+\fr{1}{1+k_0}-1 \right]
\label{F3} \ea

\noi This upper bound is worth comparing with the lower bounds to
the piezoelectric polaron GSE obtained in \cite{lbTW} for the case
of small

\bg E_{LBS}\approx -(2\alpha/\pi)\ln(k_0+1),\qquad \alpha\ll 1,
\label{small}\en

\noi and large

\bg E_{LBL}\approx -\fr{1}{3}\alpha^2
-(4\alpha/\pi)\ln(k_0/\alpha),\qquad 1\ll \alpha\ll k_0
\label{large}\en

\noi coupling constant respectively. The bound $E_{LBS}$
coincides, actually, with the perturbation theory result
\cite{pert}. It was assumed in \cite{lbTW} that $k_0\approx 150$.
So, the bounds (\ref{LLP}), (\ref{small}) and (\ref{large}) are
plotted in Figs.1-3 just for this value of $k_0$.

\section{Improvable upper bounds to the GSE of the slow-moving piezoelectric polaron}

In general case $\vP \ne 0$

\ba E_g(\alpha,\vP,k_0)\le \langle 0|{\mathcal H}(f)|0\rangle
=P^2+2\dsum_{\vk}V_kf_{\vk}+\dsum_{\vk}(k+k^2)f^2_{\vk}
-2\dsum_{\vk}(\vP\cdot {\bf k} )f^2_{\vk}+(\dsum_{\vk}
f^2_{\vk}\vk)^2,
 \label{gencase}\nn\ea

\noi with the right-hand side to be minimized by

\[ f_{\vk}=-V_k/[k-2\vk\cdot\vP(1-\eta)+k^2], \]

\noi where $\eta$ is defined self-consistently by the equation

\[ \eta\vP=
\dsum_{\vk}f^2_{\vk}\vk=\dsum_{\vk}V_k^2\vk/[k-2\vk\cdot\vP(1-\eta)+k^2]^2,\]

\noi or, alternatively, by

\[ \eta P^2
=\dsum_{\vk}V_k^2\vk\cdot\vP/[k-2\vk\cdot\vP(1-\eta)+k^2]^2.\]

\noi  The resulting upper bound is

\ba E_g(\alpha,\vP,k_0)\le P^2(1-\eta)^2- \dsum_{\vk}V_k^2
\fr{k+k^2-4\vk\cdot\vP(1-\eta)}{[k-2\vk\cdot\vP(1-\eta)+k^2]^2}.
 \label{resbound}\ea

\noi
 Another choice

\ba
f_{\vk}=-[V_k+2\eta\vk\cdot\vP]/[k-2\vk\cdot\vP+k^2],\label{compchoice}
\ea

\noi eliminating all terms linear in  Bose operators $\aka$, $\ak$
in (\ref{finalHam}), is possible too, with the corresponding
self-consistency equation for $\eta$

\[ \eta P^2=
\dsum_{\vk}f^2_{\vk}\vk=\dsum_{\vk}\vk\cdot\vP[V_k+2\eta\vk\cdot\vP
]^2/[k-2\vk\cdot\vP+k^2]^2,\]

\noi which can be solved analytically. At the same time, the
simplest choice

 \[
f_k=-V_k/(k+k^2) \]

\noi seems to be  preferable, because in this case the complexity
of the analytical calculation of arbitrary moments $\langle
0|{\mathcal H}^m(f)|0\rangle$ does not exceed the one for the case
$\vP=0$, i.e. no employment of any integrations over wave vectors
more complicated and laborious than the integrals of the type
(\ref{exint}) is necessary.

\section{Summary}

 It was shown that the GSE  function $E_g(\alpha,\vP,k_0)$
 of the slow-moving piezoelectric polaron can be approximated
  by infinite  sequence of non-increasing upper bounds for arbitrary
 values of the coupling constant $\alpha$,  polaron total momentum
$\vP$ and cut-off wave vector $k_0$.  The proposed algorithm for
the construction of these bounds is well-suited for implementation
by means of contemporary techniques for parallel computing due to
its overwhelming reliance on the Wick theorem.


\begin{wrapfigure}{i}{0.5\textwidth}
\centerline{\includegraphics[width=0.46\textwidth]{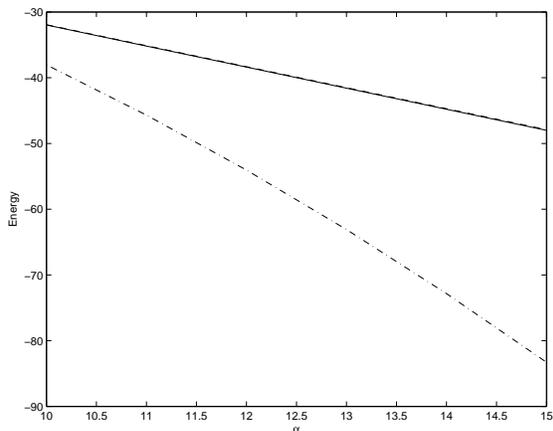}}
\caption{Upper bounds:  $E_{var}$, solid line; $E_{LBS}$, dashed
line; $E_{LBL}$, dash-dotted line; $k_0=150$. } \label{fig1}
\end{wrapfigure}

\begin{wrapfigure}{i}{0.5\textwidth}
\centerline{\includegraphics[width=0.46\textwidth]{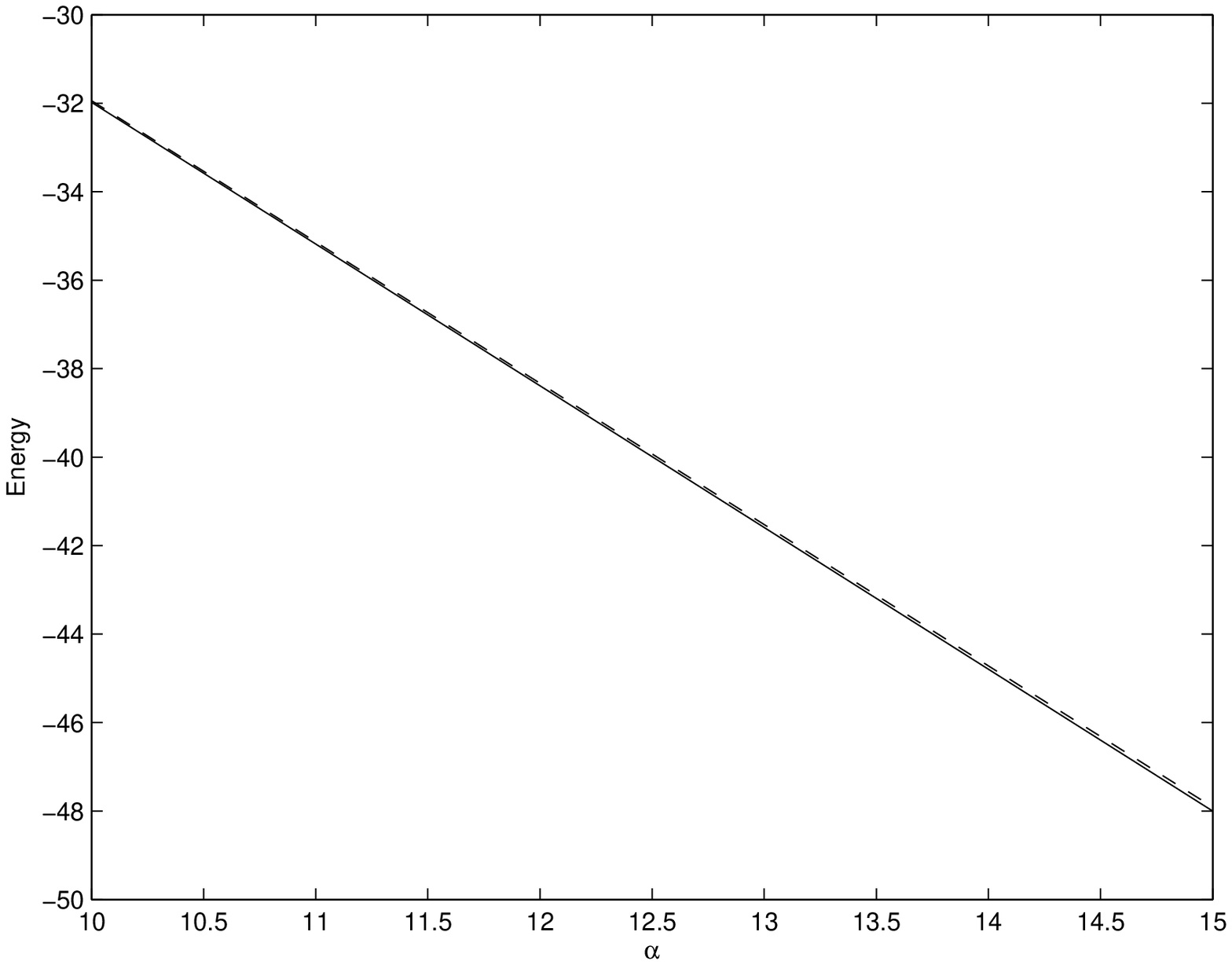}}
\caption{Upper bounds:  $E_{var}$, solid line; $E_{LBS}$, dashed
line; $k_0=150$. } \label{fig2}
\end{wrapfigure}

\begin{wrapfigure}{i}{0.5\textwidth}
\centerline{\includegraphics[width=0.46\textwidth]{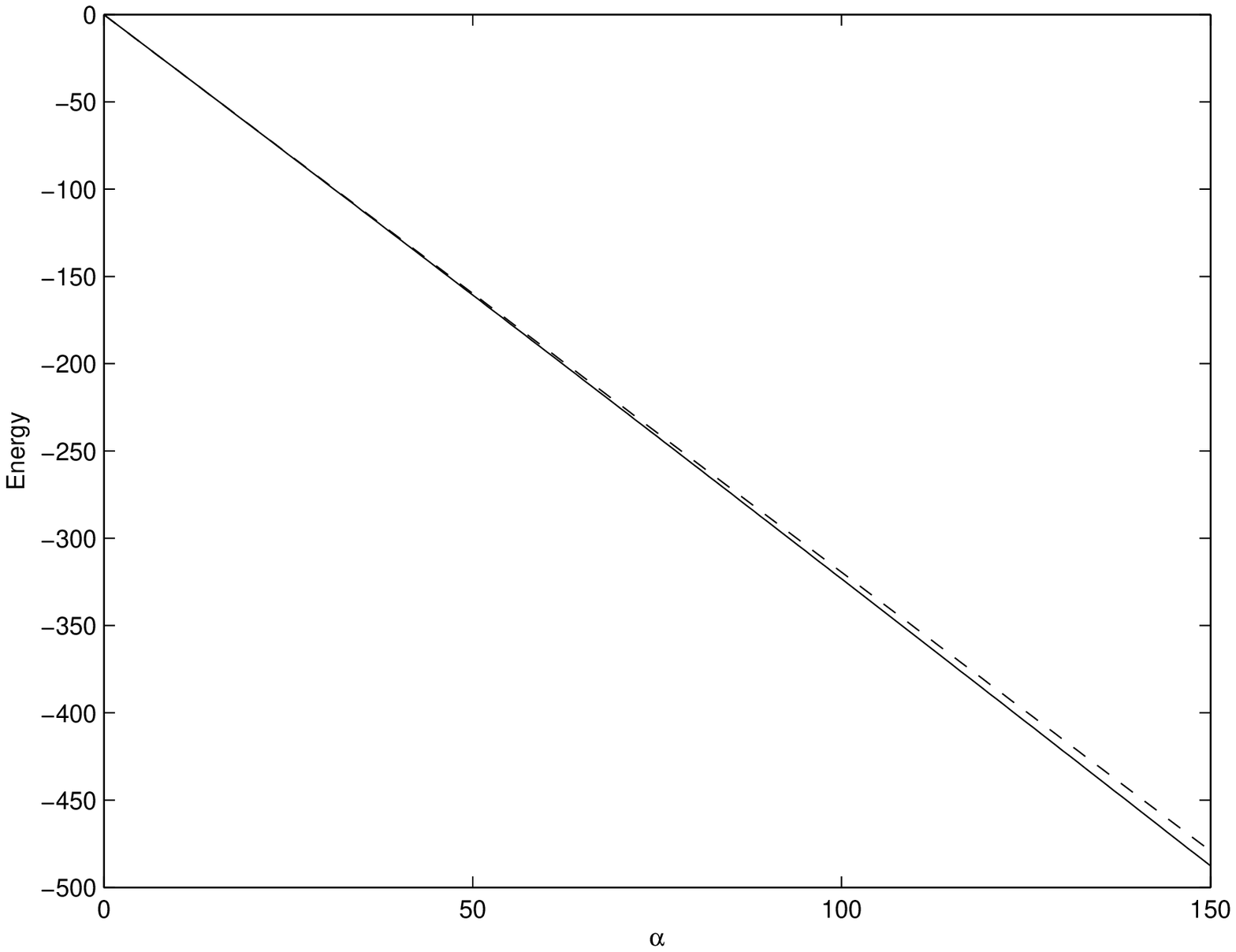}}
\caption{Upper bounds:  $E_{var}$, solid line; $E_{LBS}$, dashed
line;  $k_0=150$. } \label{fig3}
\end{wrapfigure}

 \label{last@page}

\begin{thebibliography}{10}


\bibitem{piezotens}
Hutson~A.R., J. Appl. Phys., 1961, \textbf{32},   2287.


\bibitem{Fr2} Fr\"ohlich~H., Proc. Roy. Soc. A, 1937,
\textbf{160}, 230.

\bibitem{Fr3} Fr\"ohlich~H., Advan. Phys., 1954,
\textbf{3}, 325.


\bibitem{SolVar1}  Soldatov~A.V.,  Int. J. Mod. Phys. B, 1995, \textbf {9}, n.22,
2899.


\bibitem{Kir1}
 Kireev~A.N., Int. J. Mod. Phys. B, 1997, \textbf{11}, n.10, 1235.

\bibitem{lbTW}
Thomchick~J., Whitfield~G., Phys. Rev. B,  1974, \textbf{9}, 1506.



\bibitem{pert}
Thomchick~J., Whitfield~G., Gerstner~J., Tharmalingam~K., Phys.
Rev., 1968, \textbf{165}, 993.






\end{thebibliography}
  \end{document}